\documentclass[preprint,12pt]{elsarticle}

\usepackage{amssymb}

\usepackage{bm}%
\expandafter\ifx\csname package@font\endcsname\relax\else
 \expandafter\expandafter
 \expandafter\usepackage
 \expandafter\expandafter
 \expandafter{\csname package@font\endcsname}%
\fi

\usepackage{amssymb}%nobibnotes,showpacs
\usepackage{amsmath}
\usepackage{color}

\journal{Physics Letter B}

\begin{document}

\begin{frontmatter}



\title{Solution of the noncanonicity puzzle in General Relativity: a new Hamiltonian formulation}


\author{Francesco Cianfrani$^1$, Matteo Lulli$^2$, Giovanni Montani$^{3}$.}

\address{$^1$Institut f\"ur Theoretische Physik III, Lehrstuhl f\"ur Quantengravitation,
Universit\"at Erlangen-N\"urnberg, Staudtstrasse 7, D-91058 Erlangen, EU, Germany \\

$^2$Dipartimento di Fisica, Universit\`a di Roma ``Sapienza'', Piazzale Aldo Moro 5, 00185
Roma, Italy.\\
$^3$ENEA, Centro Ricerche Frascati, U.T. Fus (Fus. Mag. Lab.), Via Enrico Fermi 45, 00044 Frascati, Roma, Italy.\\
}

\begin{abstract}
We study the transformation leading from Arnowitt, Deser, Misner (ADM) Hamiltonian formulation of General Relativity (GR) to the $\Gamma\Gamma$ metric Hamiltonian formulation derived from the Lagrangian density which was firstly proposed by Einstein. We classify this transformation as \emph{gauged} canonical - \emph{i.e.} canonical modulo a gauge transformation. In such a study we introduce a new Hamiltonian formulation written in ADM variables which differs from the usual ADM formulation mainly in a boundary term firstly proposed by Dirac. Performing the canonical quantization procedure we introduce a new functional phase which contains an explicit dependence on the fields characterizing the $3+1$ splitting. Given a specific regularization procedure our new formulation privileges the symmetric operator ordering in order to: have a consistent quantization procedure, avoid anomalies in constraints algebra, be equivalent to the Wheeler-DeWitt (WDW) quantization.      
Furthermore we show that this result is consistent with a path-integral approach. 
\end{abstract}

\begin{keyword}
Hamiltonian formulation of Gravity, Quantum Gravity.
\end{keyword}

%PACS: 04.20.Fy, 04.60.Ds.

\end{frontmatter}

\section{Introduction}
The attempts towards the quantization of GR can be classified as canonical or covariant. The latter are based on a path-integral formulation (as Causal Dynamical Triangulation \cite{cdt}, Spin-Foam models \cite{SF} and the Asymptotic Safety scenario \cite{ASG}), while the former address a canonical quantization procedure by promoting phase space coordinates to quantum operators and imposing the associated constraints \emph{\`a la} Dirac. In particular, quantum geometrodynamics (see \cite{gd} for a recent review) is based on the ADM Hamiltonian formulation \cite{ADM1}. This formulation exploits the symmetries of gravity in a $3+1$ representation, introducing new variables instead of the metric ones, where the most general set of coordinate transformations is reduced to arbitrary 3-dimensional transformations and time reparametrizations. In view of these symmetries, the configuration space can be described in Superspace, {\it i.e.} the space of Riemannian metric modulo diffeomorphisms, by requiring the vanishing of the Superhamiltonian $\mathcal{H}$ operator. On a quantum level, states are functional of 3-geometries and $\mathcal{H}=0$ is translated into the WDW equation \cite{deW}. The regularization can be performed via the heat kernel expansion \cite{KOW}. The main issues of this approach towards Quantum Gravity are the fixing of a proper operator ordering \cite{k86} and the definition of a suitable time variable \cite{k81}, which would enable to infer a conserved scalar product among physical states. 

Loop Quantum Gravity (LQG) \cite{Th,REV} is an alternative canonical formulation, in which the configuration space is parametrized by some SU(2) connections  \cite{Asht} related with ADM variables by a canonical transformation. Within this scheme, the kinematical Hilbert space is defined from the space of distributional connections. Furthermore, the Superhamiltonian operator is regularized \cite{qsd}, but the complexity of its action does not allow us to solve it analytically and physical space is still out of control.   

Henceforth, quantum geometrodynamics and LQG are both based on the ADM Hamiltonian formulation. Indeed, the Lagragian formulation adopted by Einstein is based on the so-called $\Gamma\Gamma$ Lagrangian, $\mathcal{L}_{\Gamma\Gamma}$, and on using metric components as configurational variables. 

In a recent work \cite{KK1} it is affirmed that ADM Hamiltonian formulation is not linked by a canonical transformation \cite{foot} to the  Hamiltonian formulation proposed by Dirac in \cite{Dirac} and recently extended in \cite{KK3}; see also \cite{TAS}. This result is taken as a sufficient condition to claim the \emph{nonequivalence} of these approaches for the classical dynamics. The same authors showed in \cite{KK2} that Dirac's formulation is canonically linked to $\Gamma\Gamma$ Hamiltonian formulation thus extending the noncanonicity of ADM formulation to the primitive metric formulation. This claim is however falsified by an early work \cite{ADM2} which shows that ADM equations of motion are equivalent to Einstein's field equations. The key point that the authors of \cite{KK1} are missing is that \emph{all} quantities in constrained systems are defined up to linear combinations of constraints.

We will show that the commonly used ADM Lagrangian density cannot be obtained from the $\Gamma\Gamma$ one through the ADM transformation of the metric tensor alone. Indeed a boundary term firstly used by Dirac \cite{Dirac} is needed. Transformations of the phase space variables induced by a boundary term or by a transformation of Lagrangian variables are expected to be canonical for unconstrained systems. We will see that it is not necessary for constrained systems.

In this respect, we shall propose a new Hamiltonian formulation of GR, taking as a starting point the $\Gamma\Gamma$ Lagrangian density written in ADM variables. The new phase space variables, with respect to the metric ones, have fundamental PB which are canonical modulo a gauge transformation - \emph{i.e.} modulo the PB action of a first class constraint on some function. We shall define this kind of canonicity as \emph{gauged}, differently from the \emph{weakly} canonical transformations analyzed in \cite{Kapt}, where the fundamental PB between two sets of phase space variables are canonical on the constraints hypersurfaces - \emph{i.e.} canonical modulo a combinations of constraints. Clearly both modifications of the notion of canonicity lead to classically equivalent dynamics. 

We shall then recover the common ADM formulation by means of Dirac's boundary term \cite{Dirac}, which implements a transformation which is canonical all over the phase space. We define this kind of transformation as \emph{strongly} canonical.

Our new Hamiltonian formulation is linked to the $\Gamma\Gamma$ Hamiltonian formulation via a \emph{gauged} canonical transformation and via a \emph{strongly} canonical transformation to the ADM Hamiltonian formulation. Hence, the ADM Hamiltonian formulation is linked via a \emph{gauged} canonical transformation to the $\Gamma\Gamma$ Hamiltonian formulation. Such a classification of the notion of canonicity fully explains the misleading conclusions reported in \cite{KK1}. 

We shall analyze the Hamilton-Jacobi formulation of the new theory finding the ADM secondary constraints as a transformation of the new ones, showing that the reduced phase spaces are symplectically isomorphic. 

Finally we shall perform the canonical quantization procedure on the new Hamiltonian formulation. A new functional dependence on the entire set of ADM variables in the wave functional is \emph{uniquely} determined. We will outline how the consistency with the quantum framework based on ADM formulation privileges the symmetric operator ordering for the WDW equation, given a specific regularization procedure \cite{KOW}.

In particular, the former property will be recognized as due to Dirac's boundary term. We shall justify this result from a path-integral point of view \cite{HH}. The preference of suitable operator orderings is an unexpected quantum effect and together with the result of gauged canonicity is the main result of this analysis. 

The manuscript is organized as follows. In section \ref{1} $\Gamma\Gamma$, Dirac and ADM Hamiltonian formulations are reviewed and the relationship between their Lagrangian densities is inferred. Section \ref{2} is devoted to define the canonical transformations linking different phase space variables, while in section \ref{3} the Hamilton-Jacobi equations for the new Hamiltonian formulation is discussed. In section \ref{4} the canonical quantization program is addressed and the emergence of a functional phase, leading to the ADM quantum framework, is outlined. Finally, in section \ref{5} brief concluding remarks follow.    

\section{Hamiltonian formulations for gravity}\label{1}
Let us begin by introducing the main features of ADM and $\Gamma\Gamma$ Hamiltonian formulations. We shall describe all properties with respect to the Einstein-Hilbert (EH) Lagrangian density, $\mathcal{L}_{\scriptscriptstyle{EH}} =\alpha \sqrt{-g} R$, where $\alpha = -(16 \pi l_p^2)^{-1} \hbar$ is a dimensional constant which we set equal to 1 in the classical analysis in which $l_p$ is the Planck length. The ADM transformation for the metric tensor reads
\begin{equation}
g_{00} = -N^2 + N^a N^b h_{ab}, \quad g_{0i} = N^a h_{ai}, \quad g_{ij} = h_{ij},
\label{ADMtransf}
\end{equation} 
where $N$ is the \emph{lapse function}, $N^i$ the \emph{shift vector} and $h_{ij}$ the induced \emph{three-metric}. Furthermore we shall use $K_{\mu\nu}$ and $K=g^{\mu\nu}K_{\mu\nu}$ to indicate the \emph{extrinsic curvature} tensor and its trace, $\bar{R}$ the three-dimensional scalar curvature and $\eta^\mu$ the normal vector to the spatial hypersurfaces. 

The most common way of defining the EH Lagrangian density in ADM variables is geometrical \cite{Th,REV} and it splits the action into two separated parts: a kinematical part, $\mathcal{L}_{\scriptscriptstyle{ADM}}=N \sqrt{h} (K_{\mu\nu} K^{\mu\nu} - K^2 + \bar{R})$, containing powers of fields temporal derivatives - \emph{i.e.} velocities - plus a boundary term, $\partial_\mu \boldsymbol{\mathcal{ADM}}^\mu = 2\partial_\mu [N\sqrt{h}(\eta^\mu K - \eta^\gamma \nabla_\gamma \eta^\mu)]$,  which happens to be covariant under four-diffeomorphisms. Because of the spatial second order derivative terms contained in $\bar{R}$ we need to fix the boundary conditions for Hamilton's variational principle to be well posed: choosing a manifold $\mathcal{M}$ with topology $\mathcal{M}:\mathbb{R}\times\Sigma_3$ we must impose $\partial \Sigma_3 = \emptyset$ so that the derivatives of $\delta	h_{ij}$ normal to the spatial boundary vanish. This is a common procedure \cite{ADM1,deW,Th} which however turns out not to be necessary in the Hamiltonian treatment because of the spatial nature of the divergence part of $\bar{R}$. Otherwise from a Lagrangian point of view it would be possible to avoid the hypothesis $\partial \Sigma_3 = \emptyset$ imposing the additional condition $\partial_{\bot}\delta	h_{ij}=0$ for the derivatives normal to the spatial hypersurfaces boundary. The primary constraints of this theory are
\begin{equation}
\pi \approx 0, \quad \pi_k \approx 0,
\end{equation} 
being the conjugate momenta to $N$ and $N^k$.  
With $\pi^{ij}$ as the conjugate momenta to $h_{ij}$ we shall use these symbols for the ADM Hamiltonian formulation. One usually dismisses the boundary term which would bring in accelerations making more complex the canonical treatment of the theory. 

On the other hand we can split the EH Lagrangian density, written in the natural metric variables, obtaining a kinematical part, the $\Gamma \Gamma$ part used for the Hamiltonian formulation in the early works of Pirani, Schild and Skinner \cite{Pirani,Pirani2}, $\mathcal{L}_{\scriptscriptstyle{\Gamma\Gamma}} = \sqrt{-g} g^{\mu\nu} (\Gamma^\sigma_{\mu\rho}\Gamma^\rho_{\nu\sigma} - \Gamma^\rho_{\mu\nu}\Gamma^\sigma_{\rho\sigma})$, containing powers of \emph{any} field derivative, plus a boundary term, $\partial_\mu \boldsymbol{\mathcal{EH}}^\mu=\partial_\mu[\sqrt{-g}(g^{\rho\sigma}\Gamma^\mu_{\rho\sigma}-g^{\mu\rho}\Gamma^\sigma_{\rho\sigma})]$, which in this case is not covariant. The primary constraints have a more complicated form
\begin{equation}
\psi^{0\mu} = p^{0\mu} - \frac{\partial\mathcal{L}_{\scriptscriptstyle{\Gamma\Gamma}}}{\partial \partial_0 g_{0\mu}} = 
p^{0\mu}-f^\mu(g_{\alpha\beta},\partial_k g_{\alpha\beta}) \approx 0,
\label{GGcons}
\end{equation}
being $p^{\mu\nu}$ the conjugate momenta to $g_{\mu\nu}$. We would like to emphasize that $\mathcal{L}_{\scriptscriptstyle{\Gamma\Gamma}}$, which was firstly proposed by Einstein, is the only Lagrangian density that leads to a well posed Hamilton variational principle \cite{HawkGibb} without making any hypothesis on the spacetime boundary. 

Because of their different transformation properties these divisions do not map onto each other under the ADM transformation even if we are always dealing with the same Lagrangian density which is clearly covariant.
Then, the difference should lay in some noncovariant boundary terms which must be added and subtracted thus not changing the global nature of the action.

On this assumption we determine these extra boundary terms by direct subtraction of the natural boundary terms of both formulations obtaining: 
$\partial_\mu \boldsymbol{\mathcal{EH}}^\mu
-\partial_\mu \boldsymbol{\mathcal{ADM}}^\mu =
\partial_\mu \boldsymbol{\mathcal{D}}^\mu +
\partial_k \boldsymbol{\mathcal{S}}^k +
\partial_k \boldsymbol{\mathcal{R}}^k$, where $\partial_\mu \boldsymbol{\mathcal{D}}^\mu$ is the boundary term used by Dirac in \cite{Dirac} to simplify the primary constraints \eqref{GGcons}, $\partial_k \boldsymbol{\mathcal{S}}^k$ and $\partial_k \boldsymbol{\mathcal{R}}^k$ are spatial boundary terms where the latter is caused by the presence of $\bar{R}$ in $\mathcal{L}_{\scriptscriptstyle{ADM}}$. These boundary terms read
\begin{equation}
\begin{split} 
\partial_\mu \boldsymbol{\mathcal{D}}^{\mu} = & \partial_0 \left(\frac{\sqrt{h}}{N} \partial_k N^k \right) - \partial_k \left(\frac{\sqrt{h}}{N} \partial_0 N^k \right),\\
\partial_k \boldsymbol{\mathcal{S}}^{k} = & \partial_k \left(
\frac{\sqrt{h} N^i}{N}\partial_i N^k - \frac{\sqrt{h} N^k}{N}\partial_i N^i \right),\\
\partial_k \boldsymbol{\mathcal{R}}^{k} = & \partial_k \left [ N \sqrt{h} h^{ij} h^{rk}
\left(\partial_i h_{jr} - \partial_r h_{ij} \right) \right].
\end{split}
\end{equation}
Checking the complementary result on the kinematical parts we obtain a new algebraic relation between $\mathcal{L}_{\scriptscriptstyle{ADM}}$ and $\mathcal{L}_{\scriptscriptstyle{\Gamma\Gamma}}$ which reads
\begin{equation}
\mathcal{L}_{\scriptscriptstyle{ADM}} = \mathcal{L}_{\scriptscriptstyle{\Gamma\Gamma}}
+ \partial_\mu \boldsymbol{\mathcal{D}}^{\mu} + \partial_k \boldsymbol{\mathcal{S}}^{k}
+ \partial_k \boldsymbol{\mathcal{R}}^{k}.
\label{LagLink}
\end{equation}
In order to discuss whether \eqref{LagLink} results in a canonical transformation between the Hamiltonian formulations of $\mathcal{L}_{\scriptscriptstyle{\Gamma\Gamma}}$ \cite{KK3} and $\mathcal{L}_{\scriptscriptstyle{ADM}}$ \cite{ADM1, REV, Th}, Dirac's formulation \cite{Dirac} is needed as an intermediate step. The Lagrangian density used by Dirac is given by
\begin{equation}
\mathcal{L}_{\scriptscriptstyle{D}} = \mathcal{L}_{\scriptscriptstyle{\Gamma\Gamma}} + \partial_\mu \boldsymbol{\mathcal{D}}^{\mu}.
\end{equation} 
For Dirac's formulation we shall use $\tilde{p}^{\mu\nu}$ for the conjugate momenta keeping in mind that the primary constraints are 
\begin{equation}
\tilde{p}^{0\mu} \approx 0,
\end{equation}
because Dirac's boundary term is chosen in order to have
\begin{equation}
\tilde{p}^{0\mu} = \frac{\partial \mathcal{L}_{\scriptscriptstyle{\Gamma\Gamma}}}{\partial\partial_0 g_{0\mu}} + \frac{\partial(\partial_\mu \boldsymbol{\mathcal{D}}^{\mu})}{\partial\partial_0 g_{0\mu}} = f^\mu - f^\mu \approx 0.
\end{equation}

\section{Transformations between different phase space coordinates}\label{2}
Now, our aim is to compose the ADM transformation on the metric tensor with the insertion of Dirac's boundary term $\partial_\mu \boldsymbol{\mathcal{D}}^{\mu}$ in order to study the canonicity of \eqref{LagLink}. We can follow two different ways: we evaluate the mapping of the conjugate momenta either by starting from the insertion of the boundary term followed by the transformation of variables or we proceed in the reverse way. We shall name the $\Gamma\Gamma$ Lagrangian density written in ADM variables as $\mathcal{L}_{\scriptscriptstyle{\Gamma\Gamma}}^*$. In the Hamiltonian formulation of $\mathcal{L}_{\scriptscriptstyle{\Gamma\Gamma}}^*$, $\mathcal{H}_{\scriptscriptstyle{\Gamma\Gamma}}^*$, we shall indicate the conjugate momenta associated to $N$, $N^k$ and $h_{ij}$ with $\Pi$, $\Pi_k$ and $\Pi^{ij}$, respectively.

We begin by performing the ADM transformation on $\mathcal{L}_{\scriptscriptstyle{\Gamma\Gamma}}$ giving rise to a new Hamiltonian formulation. At the Lagrangian level if we perform a direct comparison of the definitions of the conjugate momenta we obtain  
\begin{equation}
\begin{split}
& \Pi^{\scriptscriptstyle{L}} =  -2Nf^0, \quad \Pi_i^{\scriptscriptstyle{L}} = 2N^j h_{ji} f^0 + 2h_{ij} f^j, \\
& \Pi^{ij}_{\scriptscriptstyle{L}}  =  N^i N^j f^0 + 2N^{(i}f^{j)}+p^{ij},
\end{split}
\label{NcanMapGR}
\end{equation} 
which is not canonical. We can however impose the canonicity of the transformation starting from the Hamiltonian formulation of $\mathcal{L}_{\scriptscriptstyle{\Gamma\Gamma}}$, $\mathcal{H}_{\scriptscriptstyle{\Gamma\Gamma}}$, expressed in metric variables, through the request
\begin{equation}
p^{\mu \nu} \delta g_{\mu \nu} = \Pi^{\scriptscriptstyle{C}} \delta N 
+ \Pi_k^{\scriptscriptstyle{C}} \delta N^k + \Pi^{ij}_{\scriptscriptstyle{C}} \delta h_{ij}
\end{equation}
which eventually leads to the canonical form of the transformation which reads
\begin{equation}
\begin{split}
& \Pi^{\scriptscriptstyle{C}} =  -2Np^{00} \approx \Pi^{\scriptscriptstyle{L}}, \quad \Pi_i^{\scriptscriptstyle{C}} = 2N^j h_{ji} p^{00} + 2h_{ij} p^{0j} \approx \Pi_i^{\scriptscriptstyle{L}}, \\
& \Pi^{ij}_{\scriptscriptstyle{C}}  =  N^i N^j p^{00} + 2N^{(i}p^{j)0}+p^{ij} \approx \Pi^{ij}_{\scriptscriptstyle{L}},
\end{split}
\label{canMapGR}
\end{equation} 
It is clear that \eqref{NcanMapGR} differs from \eqref{canMapGR} in combinations of first class constraints, then we can state that the transformation \eqref{NcanMapGR}, which is due to the ADM metric transformation \eqref{ADMtransf} acting on $\mathcal{L}_{\scriptscriptstyle{\Gamma\Gamma}}$, is \emph{gauged} canonical; for instance the fundamental PB:
\begin{equation}
[N,\Pi^{\scriptscriptstyle{L}}]_{g,p} = [N,\Pi^{\scriptscriptstyle{C}}]_{g,p} + 2N[N,\psi^{00}]_{g,p} = 1 + 2N[N,\psi^{00}]_{g,p} = 0.
\end{equation}
This relation explicitly shows that the canonical result is altered by the PB action of a first class constraint - \emph{i.e.} a gauge transformation - thus explaining the noncanonicity result $[N,\Pi^{\scriptscriptstyle{L}}]_{g,p}=[N,f^0]_{g,p}=0$.

We define then two Hamiltonian densities: the one calculated from the transformation of $\mathcal{L}_{\scriptscriptstyle{\Gamma\Gamma}}$, and the one obtained imposing the canonicity on the ADM transformation of variables performed on $\mathcal{H}_{\scriptscriptstyle{\Gamma\Gamma}}$. These two Hamiltonian densities will differ in combinations of constraints, which are all first class \cite{KK3}, so the equations of motion will differ in a gauge transformation. The use of $\mathcal{H}_{\scriptscriptstyle{\Gamma\Gamma}}^*$ is then fully justified and we shall denote its primary constraints with $\phi$ and $\phi_k$.

We continue now with the insertion of the boundary term which links $\mathcal{L}_{\scriptscriptstyle{\Gamma\Gamma}}^*$ to $\mathcal{L}_{\scriptscriptstyle{ADM}}$. In this case the evaluation of the transformation on the conjugate momenta performed at the Lagrangian level \cite{KK2} coincides with the one performed at the Hamiltonian level which is given by the relation
\begin{equation}
\begin{split}
& \Pi \partial_0 N + \Pi_k \partial_0 N^k + \Pi^{ij} \partial_0 h_{ij} + \partial_\mu \boldsymbol{\mathcal{D}}^\mu  \\
= & \pi \partial_0 N + \pi_k \partial_0 N^k + \pi^{ij} \partial_0 h_{ij}  
\end{split}
\end{equation}
and reads
\begin{equation}
\begin{split}
\pi = \phi \approx 0, \quad \pi_k = \phi_k \approx 0, \quad \pi^{ij} = \Pi^{ij} + \frac{\sqrt{h}}{2N} h^{ij} 
\partial_k N^k.
\end{split}
\label{ADMbound}
\end{equation}
We notice that Dirac's boundary term plays the same role with both metric and ADM variables: it simplifies the primary constraints. This transformation is canonical everywhere in the phase space, hence \emph{strongly} canonical, differently from \eqref{NcanMapGR}. The two remaining boundary terms will not change this result. Hence transformation \eqref{LagLink} is \emph{gauged} canonical. 

We can discuss now the other procedure. Again, the transformation induced by Dirac's boundary term in the metric formulation, from $\mathcal{L}_{\scriptscriptstyle{\Gamma\Gamma}}$ to $\mathcal{L}_{\scriptscriptstyle{D}}$, is \emph{strongly} canonical \cite{KK2}. Performing the ADM transformation of variables on the Lagrangian density $\mathcal{L}_{\scriptscriptstyle{D}}$ we obtain the result of \cite{KK1}: the conjugate momenta to $h_{ij}=g_{ij}$ have the same definition and we do not know how to link the primary constraints properly. Thus we write, exploiting the main freedom of constrained systems
\begin{equation}
\pi^{\scriptscriptstyle{L}} = \mathcal{A}_\mu \tilde{p}^{0\mu} \approx 0, \quad 
\pi_k^{\scriptscriptstyle{L}} = \mathcal{B}_{k\mu} \tilde{p}^{0\mu} \approx 0, \quad
\pi_{\scriptscriptstyle{L}}^{ij} = \tilde{p}^{ij} + \mathcal{C}^{ij}_\mu \tilde{p}^{0\mu}.
\end{equation} 
Of course we can always fix the arbitrary coefficients $\mathcal{A}_\mu$, $\mathcal{B}_{k\mu}$ and $\mathcal{C}^{ij}_\mu$, and reproduce the canonical form of the transformation which formally coincides with \eqref{canMapGR}. Then this transformation is \emph{gauged} canonical. The insertion of the two residual spatial boundary terms will not affect this result.

The ADM Hamiltonian formulation is \emph{gauged} canonically related with the $\Gamma\Gamma$ Hamiltonian formulation. The apparent noncanonicity is now explained. For constrained systems a canonical transformation can be classified as \emph{strong}, \emph{gauged} or \emph{weak} \cite{Kapt}: the first type coincides with the definition of canonicity of unconstrained systems, while the other two are peculiar features of constrained systems. A \emph{gauged} canonical transformation implies a gauge transformation on the dynamics.

At the same time we propose a new Hamiltonian formulation, starting from the $\Gamma\Gamma$ Lagrangian but adopting the  ADM variables. This formulation is \emph{gauged} canonically related to the $\Gamma\Gamma$ Hamiltonian formulation and \emph{strongly} canonically related to the ADM one. Therefore we can relate the origin of the gauged canonicity to the use of ADM variables which implement a diffeomorphism transformation on the original metric ones.

\section{Hamilton-Jacobi equations}\label{3}
We continue our analysis with the Hamilton-Jacobi (HJ) equations for the new Hamiltonian formulation of $\mathcal{L}^*_{\scriptscriptstyle{\Gamma\Gamma}}$, which exploits the great simplification due to ADM variables. The constraints read
\begin{equation}
\begin{split}
\phi =  \Pi & - \frac{\sqrt{h}}{N^2}\partial_k N^k \approx 0, \quad \phi_k = \Pi_k -\partial_k 
\left(\frac{\sqrt{h}}{N}\right) \approx 0, \\
& \chi_i = \mathcal{H}_i + \sqrt{h} \partial_i \left(\frac{1}{N}\partial_k N^k \right) \approx 0, \\
\chi  = -&\mathcal{H} + \frac{3\sqrt{h}}{8N^2}\partial_i N^i \partial_k N^k + \frac{1}{2N}\Pi^{rs} h_{rs} \partial_k N^k \approx 0,
\end{split}\label{nc}
\end{equation}
where $\mathcal{H}=\Pi^{ab}\Pi^{ij}\mathcal{G}_{abij} -\sqrt{h}\bar{R}$ and $\mathcal{H}_i = 2h_{ij}D_a\Pi^{aj}$. Inserting ADM conjugate momenta in these quantities, instead of the new ones, we recognize $\mathcal{H}$ as the Superhamiltonian and $\mathcal{H}_i$ as the Supermomentum of the usual ADM formulation; $\mathcal{G}_{abij}=(2\sqrt{h})^{-1} (h_{ai}h_{bj}+h_{aj}h_{bi}-h_{ab}h_{ij})$ is the supermetric. The symbol $D_i$ represents an algebraic expression which has the same form of a spatial covariant derivative applied to a spatial tensor density of weight $1/2$ contracted on one index. We write the total Hamiltonian density as $\mathcal{H}^{\scriptscriptstyle{T}*}_{\scriptscriptstyle{\Gamma\Gamma}} = 
\lambda \phi + \lambda^k \phi_k + \mathcal{H}^{\scriptscriptstyle{C}*}_{\scriptscriptstyle{\Gamma\Gamma}}$ where $\lambda$ and $\lambda^k$ are Lagrange multipliers and
\begin{equation}
\begin{split}
\mathcal{H}^{\scriptscriptstyle{C}*}_{\scriptscriptstyle{\Gamma\Gamma}} = 
-N\chi & -N^i \chi_i + \partial_k \boldsymbol{\mathcal{R}}^k \\ &+ \partial_k \left( 
2\Pi^{ki}h_{ij}N^j+\frac{\sqrt{h}}{N}N^i \partial_i N^k \right),
\end{split}
\end{equation}
is known as the canonical Hamiltonian density. We notice how the absence of a spatial boundary is crucial in order to obtain $\mathcal{H}^{\scriptscriptstyle{C}*}_{\scriptscriptstyle{\Gamma\Gamma}}$ as a combination of secondary constraints only giving rise to the issue of the \emph{frozen formalism} in the canonical quantization programme. This result, for the decomposition of the Hamiltonian density, is equivalent to the one obtained in the metric formulation in \cite{KK3}. 
Now, let $S=S[N,N^k,h_{ij}]$ be Hamilton's principal functional. The request on $S$ to satisfy the primary constraints results in the decomposition 
$S=S_{\scriptscriptstyle{A}}[N,N^k,h_{ij}] + S_{\scriptscriptstyle{B}}[h_{ij}]$ where
\begin{equation}
S_{\scriptscriptstyle{A}}[N,N^k,h_{ij}] = -\int d^3x \frac{\sqrt{h}}{N}\partial_k N^k,
\end{equation}
and $S_{\scriptscriptstyle{B}}[h_{ij}]$ is not determined. Imposing the secondary constraints we have
\begin{equation}
\mathcal{H}_k\left(h_{ij},\frac{\delta S_{\scriptscriptstyle{B}}}{\delta h_{ij}}\right) \approx 0, \quad
\mathcal{H}\left(h_{ij},\frac{\delta S_{\scriptscriptstyle{B}}}{\delta h_{ij}}\right) \approx 0.
\end{equation}
Comparing with \eqref{ADMbound} it is easy to check that $\delta S_{\scriptscriptstyle{B}}/\delta h_{ij}=\pi^{ij}$. Hence, the secondary constraints of $\mathcal{H}^*_{\scriptscriptstyle{\Gamma\Gamma}}$ reduce to the ADM ones when imposed on $S$. The constraints of the ADM formulation coincide with those of the new formulation once \eqref{ADMbound} is applied. We can then state that the new constraints are all first class.
The reduced phase spaces are then symplectically isomorphic, being the hypersurfaces of constraints the same in both formulations.

The Supermomentum constraint leads to fix the dependence of $S_{\scriptscriptstyle{B}}$ on an equivalence class of three-metrics linked by a spatial diffeomorphism. We indicate this by writing $S_{\scriptscriptstyle{B}}[\{h_{ij} \}]$. The Superhamiltonian constraint imposes $S_{\scriptscriptstyle{B}}$ to be invariant under regular reparametrizations of $x^0$.

We have shown that the transformation leading from our new Hamiltonian formulation to the usual ADM one is \emph{strongly} canonical and that the constraints hypersurfaces coincide. Looking straight to \eqref{ADMbound} and \eqref{nc} and using the canonicity of the transformation we can write the equations of motion of the fundamental variables of the new formulation in terms of the ADM ones
\begin{equation}
\begin{split}
[\Pi,\mathcal{H}^{\scriptscriptstyle{T}*}_{\scriptscriptstyle{\Gamma\Gamma}}] &= \mathcal{H}^{\scriptscriptstyle{ADM}} + \left[\frac{\sqrt{h}}{N^2}\partial_k N^k, \mathcal{H}^{\scriptscriptstyle{T}}_{\scriptscriptstyle{ADM}}\right], \\
[\Pi_k,\mathcal{H}^{\scriptscriptstyle{T}*}_{\scriptscriptstyle{\Gamma\Gamma}}] &= \mathcal{H}^{\scriptscriptstyle{ADM}}_k + \left[\partial_k 
\left(\frac{\sqrt{h}}{N}\right), \mathcal{H}^{\scriptscriptstyle{T}}_{\scriptscriptstyle{ADM}}\right], \\
[\Pi^{ij},\mathcal{H}^{\scriptscriptstyle{T}*}_{\scriptscriptstyle{\Gamma\Gamma}}] &= [\pi^{ij}, \mathcal{H}^{\scriptscriptstyle{T}}_{\scriptscriptstyle{ADM}}] - \left[\frac{\sqrt{h}}{2N} h^{ij} 
\partial_k N^k, \mathcal{H}^{\scriptscriptstyle{T}}_{\scriptscriptstyle{ADM}}\right],
\end{split}
\end{equation}
where $\mathcal{H}^{\scriptscriptstyle{T}}_{\scriptscriptstyle{ADM}}$ is the total Hamiltonian density of the ADM formulation.
The transformation acts only on the conjugate momenta hence the equations of motion for $h_{ij},N^i$ and $N$ are unchanged. Furthermore because of the equivalence of the constraints hypersurfaces and the canonicity of the transformation the algebra described by the new constraints will match that of ADM formulation. It follows that the new constraints are all first class.
In the ADM formulation both $N$ and $N^i$ have weakly vanishing conjugate momenta reflecting their arbitrarity. In the new Hamiltonian formulation we propose this does not happen and a clear answer about their arbitrarity can be found in their equations of motion
\begin{equation}
[N,\mathcal{H}^{\scriptscriptstyle{T}*}_{\scriptscriptstyle{\Gamma\Gamma}}] = \lambda, \qquad 
[N^k,\mathcal{H}^{\scriptscriptstyle{T}*}_{\scriptscriptstyle{\Gamma\Gamma}}] = \lambda^k,
\end{equation}
which then result as arbitrary functions showing no link to their conjugate momenta. Hence $N$ and $N^i$ are still arbitrary. We briefly comment upon the main feature of this new Hamiltonian formulation: it has been formulated directly from the $\Gamma\Gamma$ Lagrangian density via algebraic manipulations only, no request of $3+1$ transformation properties has been imposed on ADM variables. The strong canonicity link with ADM formulation ensures the classical equivalence of these two formulations and indeed the latter has a more concise expression. However through this new Hamiltonian formulation a clear algebraic link with the $\Gamma\Gamma$ Lagrangian density is established giving some detailed information, such as the gauged canonicity link to the latter or the role of Dirac's boundary term in simplifying ADM constraints, which were hidden behind the geometrical interpretation of ADM phase space variables. We shall see in the next section that this new Hamiltonian formulation leads to some interesting consequences on the quantum level.

\section{Canonical Quantization}\label{4}
Let us now compare the quantum formulations associated with the $\mathcal{L}^*_{\scriptscriptstyle{\Gamma\Gamma}}$ and $\mathcal{L}^*_{\scriptscriptstyle{ADM}}$ (we will restore the constant $\alpha$). It is well known that in the ADM formulation the Hamiltonian is given by a combination of the secondary constraints $H=\int d^3x(N\mathcal{H}+N^k \mathcal{H}_k)$. The canonical quantization programme develops by promoting the fields as multiplicative operators and their conjugate momenta as functional derivatives times $-i\hbar$. The information is encoded into a functional of the fields $\Phi[N,N^i,h_{ij}]$ which describes the physical states once satisfied all the constraints. In the ADM formulation one gets for the primary constraints $\pi\Phi=-i\hbar \delta\Phi/\delta N=0$ and $\pi_k\Phi=-i\hbar \delta\Phi/\delta N^k=0$. These equations can be solved by a functional of $h_{ij}$ solely. Solving the Supermomentum constraint $\mathcal{H}_k \Phi[h_{ij}] = 0$ one has that the functional must depend on an equivalence class of three-metrics just like observed for the HJ treatment of the classical theory: thus we write $\Phi[\{h_{ij}\}]$. The request $\mathcal{H}\Phi =0$ leads to the WDW equation and its solution is one of the main tasks of the canonical quantization programme. Furthermore this equation needs to be somehow regularized \cite{KOW}. 

We perform now the canonical quantization on the new Hamiltonian system described by $\mathcal{H}^*_{\scriptscriptstyle{\Gamma\Gamma}}$ adopting the same space of states and requiring the vanishing of the constraints defined in \eqref{nc} once applied on the wave functional.

Let us begin by imposing the primary constraint $\phi$ on a generic wave functional $\Psi=\Psi[N,N^k,h_{ij}]$. We obtain
\begin{equation}
\begin{split}
\phi \Psi &= -i\hbar \frac{\delta \Psi}{\delta N} - \frac{\alpha \sqrt{h}}{N^2}\partial_k N^k \Psi = 0, \\ \Rightarrow &\quad \Psi[N,N^k,h_{ij}] = \exp \left \{-\frac{i\alpha}{\hbar} \int d^3x \frac{\sqrt{h}}{N}\partial_k N^k \right \}\Phi[N^k, h_{ij}].
\end{split}\label{phi}
\end{equation}
We continue now by imposing the other primary constraint, $\phi_k$
\begin{equation}
\phi_k \Psi = 0 \quad \Rightarrow \quad  \Psi[N,N^k,h_{ij}] = \exp \left \{-\frac{i\alpha}{\hbar} \int d^3x \frac{\sqrt{h}}{N}\partial_k N^k \right \}\Phi[h_{ij}],\label{phik}
\end{equation}
hence the only effect is to reduce the dependence of $\Phi$ on the three-metric. Imposing the secondary constraint $\chi_k$ we obtain
\begin{equation}
\Psi[N,N^k,h_{ij}] = \exp \left \{-\frac{i\alpha}{\hbar} \int d^3x \frac{\sqrt{h}}{N}\partial_k N^k \right \}\Phi[\{h_{ij}\}].
\label{q}
\end{equation}
The imposition of the constraints, $\phi, \phi_k$ and $\chi_k$, implies a factorization of the wave functional $\Psi = \Xi \Phi$ where $\Xi$ denotes the functional phase. This result does not depend on the choice of the operator ordering: $\phi$ and $\phi_k$ are free from ordering ambiguities; $\chi_k$ is just a modified version of the ADM Supermomentum whose action is proved to be independent on the ordering choice (see K. Kuchar's section in \cite{Israel}). Hence the functional phase is an intrinsic property of our new Hamiltonian formulation. Moreover this decomposition is the quantum analogue of the HJ analysis we just performed, hence we are led to recognize in $\Phi$ the usual ADM wave function. Hence the imposition of three of our new constraints on $\Psi$ implies the imposition of three ADM constraints on $\Phi$
\begin{equation}
\phi \Psi = \Xi (\pi \Phi) = 0, \qquad \phi_k \Psi = \Xi (\pi_k \Phi) = 0, \qquad \chi_k \Psi = \Xi (\mathcal{H}_k \Phi) = 0.
\end{equation}

\subsection{Quantum Dynamics}
As soon as $\chi$ (\ref{nc}) is promoted to be a quantum operator, two main issues arise: the need to regularize the term with two momenta and the operator ordering. In particular, the former can be addressed by introducing a proper point-splitting procedure parametrized by $t$, such that one replaces
\begin{equation}
\mathcal{G}_{ijkl}(x)\Pi^{ij}(x)\Pi^{kl}(x)\rightarrow \int d^3z G_{ijkl}(x)K(x,z;t)\Pi^{ij}(x)\Pi^{kl}(z),
\end{equation}
where the function $K(x,z;t)$ can be defined via the heat kernel expansion as shown in \cite{KOW}. From here on we will use $G_{ijab} = \sqrt{h}\mathcal{G}_{ijab}$, because of the definition of $K$
\begin{equation}
\lim_{t\to0} K(x,z;t) = \frac{\delta^{(3)}(x-z)}{\sqrt{h(x)}}.
\end{equation}
The operator ordering issue is due to the fact that two terms are generically ambiguous on a quantum level: the one containing the supermetric, which is not polynomial in the configuration variables, and the additional contribution proportional to $\Pi^{rs} h_{rs}$ which is specific of the new Hamiltonian formulation. Indeed, the ordering of these two terms is not completely arbitrary if we require $\Psi$ (\ref{q}) to be a proper solution of the whole set of constraints. In fact, as soon as the factorization of the wave function is performed, $\chi\Psi[N,N^i,h_{ij}]=0$ implies a certain condition $\chi'\Phi[\{h_{ij}\}]=0$, which can be consistently solved only if it contains no dependence on the lapse function and the shift vector. 

Let us consider the following expression for $\chi$ in which the factor-ordering choices, labeled by two real parameters $\beta$ and $\gamma$ ($\beta,\gamma\in[0,1]$), are fixed such that like in \cite{KOW} delta functions of the form $\delta^3(0)$ do not show up, 
\begin{equation}
\begin{split}\chi(\beta,\gamma)\Psi= & -\frac{1}{\alpha}\int d^{3}zG_{ijab}\left(z\right)K\left(x,z;t\right)\Pi^{ij}\left(x\right)\Pi^{ab}\left(z\right)\Psi\\
- & \frac{1-\beta}{\alpha}\int d^{3}z\left\{\Pi^{ij}\left(x\right)\left[G_{ijab}\left(z\right)K\left(x,z;t\right)\right]\right\}\Pi^{ab}\left(z\right)\Psi\\
+ & \frac{1}{2N\left(x\right)}\partial_{k}N^{k}\left(x\right)h_{ij}\left(x\right)\Pi^{ij}\left(x\right)\Psi\\
+ & \frac{1-\gamma}{4N\left(x\right)}\partial_{k}N^{k}\left(x\right)\Psi\int d^{3}z\left\{\Pi^{ij}\left(x\right)\left[h_{ij}\left(z\right)\sqrt{h\left(z\right)}K\left(x,z;t\right)\right]\right\}\\
+ & \alpha\sqrt{h\left(x\right)}\bar{R}\left(x\right)\Psi+\frac{3\alpha\sqrt{h\left(x\right)}}{8N^{2}\left(x\right)}\partial_{i}N^{i}\left(x\right)\partial_{k}N^{k}\left(x\right)\Psi,
\end{split}
\end{equation}
where in the 2nd and 4th terms $\Pi^{ij}(x)$ act only on the functions within square brackets $[\ldots]$. The heat kernel appears again in the 4th term in order to ensure that a $\delta^3(0)$ term does not show up in the final expression. 

The symmetric ordering and the one with momenta on the right correspond to $\beta=\gamma=0$ and $\beta=\gamma=1$, respectively. It is possible to consider more generic orderings, but we regard the introduction of further ambiguities in the term proportional to $\Pi^{rs} h_{rs}$ as rather unnatural (such a term is polynomial in $\Pi^{rs}$ and $h_{rs}$). The evaluation of the action of $\chi$ on $\Psi$ gives
\begin{equation}
\begin{split}\chi\left(\beta,\gamma\right)\Psi= & -\Xi\mathcal{H}\left(\beta,\gamma\right)\Phi+i\hbar\frac{\sqrt{h\left(x\right)}}{4N\left(x\right)}\partial_{k}N^{k}\left(x\right)\Psi\times\\
\times & \left[\frac{3}{2}\left(5\gamma+2\beta\right)K\left(x,x;t\right)+\left(\gamma-\beta\right)h_{ab}\left(x\right)k^{ab}\left(x,x;t\right)\right]=0,
\end{split}\label{hchi}
\end{equation}
where $\Pi^{ij}K=-i\hbar k^{ij}$. It is worth noting how in order to have a solution of the secondary constraint $\chi$ (\ref{hchi}) consistent with those already solved (\ref{phi}), (\ref{phik}) the following condition must hold
\begin{equation}
\Gamma(x;\beta,\gamma) = \frac{3}{2}\left(5\gamma+2\beta\right)K\left(x,x;t\right)+\left(\gamma-\beta\right)h_{ab}\left(x\right)k^{ab}\left(x,x;t\right) = 0.
\label{rel}
\end{equation}
The same condition ensures that the algebra of $\phi,\phi_k$ with $\chi$ is not anomalous. In fact, one finds
\begin{equation}
\begin{split}
[\phi[A], \chi[B]] &= -A(x)B(x)\frac{\hbar^2 \sqrt{h(x)}}{4 N^2(x)} \partial_k N^k(x) \Gamma(x;\beta,\gamma), \\
[\phi_k[A^k], \chi[B]] &= \partial_{k}A^{k}\left(x\right)B\left(x\right)\frac{\hbar^{2}\sqrt{h\left(x\right)}}{4N\left(x\right)} \Gamma(x;\beta,\gamma),
\end{split}
\end{equation}
$A$ and $B$ being smearing functions. Such a nonanomalous behavior allows us to use the explicit solution of $\phi \Psi=\phi_k \Psi=0$ to parametrize the hypersurfaces $\chi \Psi = 0$. Once equation \eqref{rel} is satisfied we obtain
\begin{equation}
\chi(\beta,\gamma) \Psi[N,N^k,h_{ij}] = -\Xi[N,N^k,h_{ij}] \mathcal{H}(\beta,\gamma) \Phi [\{h_{ij}\}] = 0,
\end{equation}
enforcing the quantum equivalence with the WDW formulation and giving the possibility to realize a nonanomalous algebra for the entire set of constraints as discussed in \cite{KOW2}.

Therefore, the relation (\ref{rel}) must be imposed in order to have a consistent solution of primary and secondary constraints. This condition is identically solved no matter the form of the regulator by fixing $\beta=\gamma=0$, {\it i.e.} the symmetric operator ordering.  This ordering is thus a privileged one within this formulation. On the contrary, when momenta in the Superhamiltonian stand on the right ($\beta= 1$), the regulator must be fixed such that the condition (\ref{rel}) holds. Even though one could fix $\Gamma(x;1,\gamma) = 0$ via a proper choice of the regulator one would not get a consistent ordering prescription for the whole operator $\chi$.

Hence, the requirement of dealing with a consistent set of constraints on a quantum level independently of the choice of the regulator selects a proper class of operator orderings for the Superhamiltonian operator. In particular, the symmetric ordering is admissible, while one with momenta on the right is ruled out.

This procedure is very similar to the one proposed in \cite{Halliwell} and it enforces the quantum dynamical equivalence of both formulations. Indeed, given the operator ordering, the imposition of the new constraints on $\Psi$ implies the imposition of ADM constraints on $\Phi$.
Therefore, the relation \eqref{q} maps the solutions of the set of constraints \eqref{nc} into the solutions of the ADM one, if equation \eqref{rel} is satisfied.

The relation between ADM and $\Gamma\Gamma$ wave functionals can be inferred also in a path-integral formulation. The Euclidean ground state wave functional associated with a 3 metric configuration $h_{ij}$ on a spatial hypersurfaces is given by \cite{HH}
\begin{equation}  
\Phi[\{h_{ij}\}] \propto \int D[g] e^{-S_{\scriptscriptstyle{ADM}}},
\end{equation}
where the integral is extended over all the 4-metric configurations $g_{\mu\nu}$ having a boundary on which the induced metric is $h_{ij}$. In particular for the $\Gamma\Gamma$ wave functional one finds, neglecting purely spatial boundary terms
\begin{equation}  
\Psi \propto \int D[g] e^{-S_{\scriptscriptstyle{\Gamma\Gamma}}}=\int D[g]
 e^{-S_{\scriptscriptstyle{ADM}}+\int d^4x\partial_\mu \boldsymbol{\mathcal{D}}^\mu}.
\end{equation} 
Being the difference between the Lagrangian densities a boundary term \eqref{LagLink}, it receives contributions only from the boundary configurations. This fact implies that $\exp\{\int d^4x \partial_\mu \boldsymbol{\mathcal{D}}^\mu\}$ comes out from the path integral and it gives a phase term in front of the ADM wave functional, whose evaluation on a spatial hypersurfaces gives the following expression
\begin{equation}  
\Psi \propto \exp \left \{\int d^3x{\frac{\sqrt{h}}{N}\partial_kN^k}\right \} \Phi[h_{ij}]
\label{pi}
\end{equation} 
The phase in \eqref{pi} is the Euclidean counterpart of the phase $\Xi$ obtained from the canonical analysis \eqref{q}. It is interesting to see how the path integral analysis we just performed, where the operator ordering is not an issue, supported the canonical frame we developed by confirming the presence of the functional phase. 

\section{Conclusions}\label{5}

This work has been motivated by the great historical importance of ADM formulation in quantizing GR. We solved the puzzle of the noncanonicity firstly indicated in \cite{KK1} with the concept of \emph{gauged} canonicity: a transformation is \emph{gauged} canonical if its fundamental PB are canonical modulo a gauge transformation. The authors of \cite{KK1} were seeking a \emph{strong} canonicity, the same of unconstrained systems. Clearly, the classical equivalence of the different formulations is untouched because their dynamics are equivalent on the hypersurfaces of constraints. The gauged canonicity results implies, however, that some attention must be paid in order to have a canonical ADM quantization equivalent to a yet unknown canonical $\Gamma\Gamma$ quantization, because fundamental PB are not preserved by \eqref{LagLink}.

We then proposed a new Hamiltonian formulation which exploits the great deal of simplification due to the use of ADM variables. We showed that usual ADM secondary constraints can be recovered as `reduced' constraints in the HJ treatment of the new Hamiltonian formulation. We performed the canonical quantization procedure obtaining a new wave functional which can be factorized in two terms: a functional phase containing $N$ and $N^i$, which is not $3+1$ covariant, and a functional which can be identified with the WDW functional. Furthermore, given a specific regularization prescription studied in \cite{KOW}, the formulation privileges a class of operator ordering, avoiding inconsistencies of the quantization procedure, assuring the quantum equivalence with the WDW quantization and avoiding anomalies in constraints algebra. In this respect the symmetric ordering fulfills all these requirements independently of the regulator choice, which is a desirable physical feature. The ordering result has relevant implications on the Universe dynamics as discussed in \cite{HP,M}. Furthermore we justified, from a path-integral point of view, the presence of the functional phase as due to a boundary contribution in the action. 

Further developments should study the behavior of this new wave functional under four diffeomorphisms; one attempt could use the class of solutions for $\Phi[\{h_{ij}\}]$ proposed in \cite{KOW}. It will be compelling to cast this formulation in Ashtekar's variables \cite{Asht} in order to study possible modifications of observables in LQG framework where a Hilbert space is properly defined. Furthermore it would be interesting to induce other canonical transformations via boundary terms in the ADM formulation and study whether the preference for a symmetric ordering is a universal property of the class of equivalent theories obtained via this procedure.

\section*{Acknowledgments}
This work has been partially developed in the framework of the CGW collaboration
(www.cgwcollaboration.it).

\end{document}